\documentclass[twocolumn,showpacs,showkeys,preprintnumbers,prl,amsmath,amssymb]{revtex4}

\usepackage{graphicx}
\usepackage{epsfig} 
\usepackage{dcolumn}
\usepackage{bm}

\begin{document}

\newcommand{\cll}{$c_{11}$ }
\newcommand{\clly}{$c_{11}$}
\newcommand{\cuu}{$c_{66}$ }
\newcommand{\cuuy}{$c_{66}$}
\newcommand{\cms}{cm$^{2}$ }
\newcommand{\cmsy}{cm$^{2}$}
\newcommand{\ab}{$\sim$ }
\newcommand{\aby}{$\sim$}
\newcommand{\tp}{$T'$ }
\newcommand{\tpy}{$T'$}
\newcommand{\tph}{$T_{p}$ }
\newcommand{\tphy}{$T_{p}$}
\newcommand{\too}{$T_{0}$ }
\newcommand{\tooy}{$T_{0}$}
\newcommand{\tauep}{$\tau_{ep}$ }
\newcommand{\tauepy}{$\tau_{ep}$}
\newcommand{\tauee}{$\tau_{ee}$ }
\newcommand{\taueey}{$\tau_{ee}$}
\newcommand{\jphi}{$j_{\phi}$ }
\newcommand{\jphiy}{$j_{\phi}$}
\newcommand{\tc}{$T_{c}$ }
\newcommand{\tcy}{$T_{c}$}
\newcommand{\hcl}{$H_{c1}$ }
\newcommand{\hcly}{$H_{c1}$}
\newcommand{\ef}{$E_{F}$ }
\newcommand{\efy}{$E_{F}$}
\newcommand{\estar}{$E^{*}$ }
\newcommand{\estary}{$E^{*}$}
\newcommand{\htc}{high-temperature superconductors } 
\newcommand{\htcy}{high-temperature superconductors}
\newcommand{\et}{{\it et al. }}
\newcommand{\ety}{{\it et al.}}
\newcommand{\be}{\begin{equation} }
\newcommand{\ene}{\end{equation}}
\newcommand{\hh}{$H$ }
\newcommand{\hhy}{$H$}
\newcommand{\hc}{$H_{c}$ }
\newcommand{\hcy}{$H_{c}$}
\newcommand{\ho}{$H_{0}$ }
\newcommand{\hoy}{$H_{0}$}
\newcommand{\jc}{$j_{c}$ }
\newcommand{\jcy}{$j_{c}$}
\newcommand{\sg}{superconducting }
\newcommand{\sgy}{superconducting}
\newcommand{\ssc}{superconductor }
\newcommand{\sscy}{superconductor}
\newcommand{\hcu}{$H_{c2}$ }
\newcommand{\hcuy}{$H_{c2}$}
\newcommand{\rfff}{$\rho_{f}$ }
\newcommand{\rfffy}{$\rho_{f}$}
\newcommand{\hcut}{$H_{c2}(T)$ }
\newcommand{\hcuty}{$H_{c2}(T)$}
\newcommand{\jd}{$j_{o}$ }
\newcommand{\jdy}{$j_{o}$}
\newcommand{\ybco}{Y$_{1}$Ba$_{2}$Cu$_{3}$O$_{7-\delta}$ }
\newcommand{\ybcoy}{Y$_{1}$Ba$_{2}$Cu$_{3}$O$_{7-\delta}$}
\newcommand{\lsco}{La$_{2-x}$Sr$_{x}$CuO$_{4}$ }
\newcommand{\lscoy}{La$_{2-x}$Sr$_{x}$CuO$_{4}$}
\newcommand{\de}{$\delta \epsilon$ }
\newcommand{\dey}{$\delta \epsilon$}
\newcommand{\nq}{$n_{q}$ }
\newcommand{\nqy}{$n_{q}$}
\newcommand{\rrhon}{$\rho_{n}$ }
\newcommand{\rrhony}{$\rho_{n}$}
\newcommand{\rrho}{{$\rho$} }
\newcommand{\rrhoy}{{$\rho$}}
\newcommand{\qp}{quasiparticle }
\newcommand{\qpy}{quasiparticle}
\newcommand{\qps}{quasiparticles }
\newcommand{\qpsy}{quasiparticles}
\newcommand{\bib}{\bibitem}
\newcommand{\ib}{{\em ibid. }}
\newcommand{\taue}{$\tau_{\epsilon}$ }
\newcommand{\tauey}{$\tau_{\epsilon}$}
\newcommand{\vstary}{$v^{*}$}
\newcommand{\vstar}{$v^{*}$ }
\newcommand{\tstar}{$T^{*}$ }
\newcommand{\tstary}{$T^{*}$}
\newcommand{\rhostar}{$\rho^{*}$ }
\newcommand{\rhostary}{$\rho^{*}$}
\newcommand{\vinf}{$v_{\infty}$ }
\newcommand{\vinfy}{$v_{\infty}$}
\newcommand{\fd}{$F_{d}$ }
\newcommand{\fdy}{$F_{d}$}
\newcommand{\fe}{$F_{e}$ }
\newcommand{\fey}{$F_{e}$}
\newcommand{\fl}{$F_{L}$ }
\newcommand{\fly}{$F_{L}$}
\newcommand{\jstar}{$j^{*}$ }
\newcommand{\jstary}{$j^{*}$}
\newcommand{\je}{$j(E)$ }
\newcommand{\jey}{$j(E)$}
\newcommand{\vphi}{$v_{\phi}$ }
\newcommand{\vphiy}{$v_{\phi}$}
\newcommand{\blo}{$B_{1}$ }
\newcommand{\bloy}{$B_{1}$}
\newcommand{\bhi}{$B_{\infty}$ }
\newcommand{\bhiy}{$B_{\infty}$}
\newcommand{\vlo}{$v_{1}$ }
\newcommand{\vloy}{$v_{1}$}
\newcommand{\bo}{$B_{o}$ }
\newcommand{\boy}{$B_{o}$}
\newcommand{\eo}{$E_{o}$ }
\newcommand{\eoy}{$E_{o}$}

\preprint{Accepted for publication in Phys. Rev. Lett. ($\sim$Sept. 2002)}

\title{Unstable flux flow due to heated electrons in Y-Ba-Cu-O films}

\author{Milind N. Kunchur} 
 \homepage{http://www.physics.sc.edu/kunchur}
 \email{kunchur@sc.edu}
\affiliation{Department of Physics and Astronomy\\
University of South Carolina, Columbia, SC 29208}

\date{Accepted August 5, 2002}

\begin{abstract}
A flux instability occurs in superconductors at low temperatures, 
where {\em ee} scattering is more rapid than {\em ep}, whereby the
dissipation significantly elevates the electronic temperature
while maintaining a thermal-like distribution function.
The reduction in condensate and rise in
resistivity produce a
non-monotonic current-voltage response.
In contrast to the Larkin-Ovchinnikov instability
where the vortex shrinks, in this scenario the vortex
expands and the quasiparticle population rises.
Measurements in \ybco agree quantitatively with
the distinct predictions of this mechanism.
\end{abstract}

\maketitle

In a type II superconductor, magnetic fields between the lower critical 
value \hcl and upper critical field \hcu 
introduce flux vortices  containing a quantum of 
flux $\Phi_{o}=h/2e$. Here we have
superconducting films in a perpendicular applied flux density $B$, with 
a transport electric current density $j$ in the plane of the film, which
exerts a Lorentz driving force 
$F_{L} = j \Phi_{o}$. 
The vortex motion generates an electric field $E=vB$ and is 
opposed by a viscous drag $\eta v$ ($\eta$ is the coefficient of
viscosity and $v$ the vortex velocity), 
so that in steady state $j \Phi_{o} =  \eta v$ and the response is
Ohmic. Larkin and Ovchinnikov
(1986) have shown that a dirty superconductor at low temperatures has a
free-flux-flow resistivity related to the normal-state value $\rho_{n}$
by \cite{lochapter}
\begin{equation} \label{bseq} 
 {\rho_{f}}/{\rho_{n}} \simeq 0.9 {B}/{H_{c2}(T)}. \end{equation}
Approximately the same result, without the precise 0.9 prefactor,
can be obtained by considering the Ohmic dissipation in
the core and temporal changes in the order parameter leading to
irreversible entropy transfer \cite{bsandtinkhamandclem}; the result is
also
valid for {\em d}-wave superconductors that are not superclean
\cite{kopninvolovik}. Eq.~\ref{bseq} is equivalent to $\eta \approx 
\Phi_{0}H_{c2}/\rho_{n}$. At low levels of $j$ and $E$, in the assumed 
dirty limit $l \ll \xi E_{F}/kT_{c}$ ($l$ is the mean free path and
$E_{F}$ is the Fermi energy),  $\eta$ is a constant
that is proportional to the order parameter $\Delta$ and 
inversely proportional to the size $\sim \xi^{2}$ (where $\xi$ is the
coherence length) of the vortex \cite{tinkhamtext}. 
(The Hall effect and transverse component of $E$ are
negligible for this discussion, as is vortex pinning because
of the large driving forces \cite{obs}.)

At high electric fields and dissipation levels sufficient to
alter the electronic distribution function and/or the electronic
temperature,
$j(E)$ becomes non-linear and can develop an unstable region ($dj/dE <
0$)
above some critical vortex velocity \vstary .
For the regime near \tcy , such an instability has been predicted by 
Larkin and Ovchinnikov (LO) \cite{LO}, and has 
been experimentally well established \cite{klein,loexpt,doett,xiao}. 
At high temperatures, the electron-phonon ({\em ep}) scattering time
\tauep
can be shorter than the 
electron-electron ({\em ee}) scattering time \taueey , preventing
internal
equilibration of the electronic system and producing a peculiar
non-thermal 
distribution function. Since the order parameter $\Delta$ is
especially sensitive to the distribution function when close to \tcy ,
even moderate values of $E$ can sufficiently distort $\Delta$ (via the
Eliashberg mechanism \cite{eliashberg}) causing a shrinkage of the
vortex 
core and a removal of quasiparticles from its vicinity. This is the
gist of the LO behavior \cite{LO,klein}. As LO themselves emphasize,
the effect is most favorable close to \tc for superconductors
with a full gap and, as shown by Bezuglyj and Shklovskij \cite{bezshkl},
is dominant for $B < B_{T}$ 
(with 
$B_{T} \sim 0.1$ T for our low-$T$ regime). 
One of the predictions of
the standard LO effect is a \vstar that is $B$ independent. This has
been
confirmed in \ybco in the high-$T$ range \cite{doett,xiao}.

This work investigates the opposite regime of $T \ll T_{c}$ and $B > 
B_{T}$, where $\Delta$ is not sensitive to small changes in the
distribution function. Furthermore 
because \tauee $<$ \tauep as $T\rightarrow 0$, the distribution 
function remains thermal like and the electronic system suffers 
mainly a temperature shift with respect to the lattice 
\cite{LO,bezshkl,bez,volotskaya}. Then instead of the standard LO
picture
described earlier, we consider a more transparent scenario
where the main effect of the dissipation is to raise the electronic
temperature, create additional 
quasiparticles, and diminish $\Delta$. The vortex expands rather than
shrinks, and the viscous drag is reduced because of a softening of 
gradients of the vortex profile rather than a removal of
quasiparticles. This sequence of events is almost opposite to the 
standard LO picture and represents a new type of unstable regime
prevalent at $T \ll T_{c}$. 
All experimental measurables can be calculated 
without ambiguity, and the predicted field dependencies and full \je 
curves fit the experimental results without any adjustable parameters. 

Previously some deviations from LO behavior at intermediate 
temperatures---such as a $B$-dependent \vstary ---were treated through
modifications to the LO effect, such as an 
intervortex spacing $l_{\phi}$ that exceeds 
the energy-relaxation length $l_{\epsilon}$
\cite{doett2} or by inclusion of thermal effects \cite{bezshkl,bez}.
Those
treatments do not apply to the present $T \ll$ \tc regime where
$l_{\epsilon} \sim \sqrt{D\tau_{ep}} \sim$ 100--1000 nm is larger than 
$l_{\phi} = 1.075 \sqrt{\Phi_{o}/B} \sim$ 10--50 nm 
($D= 3 \times 10^{-4}$ m$^{2}$/s is the diffusion constant
\cite{doett}).

Here we take a ``bottom-up'' approach and start from the 
$T \sim 0$ limit: The total input power $jE$ travels from 
electrons to lattice and from there to the bath, 
so that \too $<$ \tph
$<$ \tpy , where \too and \tph are the bath and phonon temperatures, and
\tp is the raised non-equilibrium electronic temperature. 
Macroscopic heating, represented by $T_{p} - T_{0} = R_{th}jE$, is 
$<5$\% of the total increase \tpy - \too for the worst case 
dissipation so that $T_{p} \approx T_{0}$ 
(Here $R_{th}$$\sim$1 nK.cm$^{3}$/W 
is the total thermal resistance between the film and the bath; 
see experimental section. It will be seen later that the specific-heat
integral heavily weights the higher temperatures, so that \tphy - \too
is quite negligible.). The energy 
relaxation between electrons
and lattice occurs by inelastic {\em ep} scattering, and is 
characterized by  an effective time \tauey $\sim$$<$\tauepy $>$. 
The principle contributions to \rrhon come from impurities and
phonons. Since the phonon temperature remains near the bath, \rrhony 's
value will not change as the non-equilibrium \tp rises.
Thus putting $\rho_{n}(T_{0})$ and 
\hcuy $(T')$ (since \hcu does depend on $T'$) into  
previous equations gives the \je response in terms of \tpy :
\begin{equation}
\label{tdep-je}
j=v \eta (T')/\Phi_{0} 
= E H_{c2}(T')/\rho_{n}(T_{0})B.
\end{equation}

One now has the ingredients for calculating the critical field
dependencies in a few steps. 
The \je function of Eq.~\ref{tdep-je} is non-monotonic since
$E$ 
is multiplied by $H_{c2}(T')$ (or $\eta(T')$) which drops 
rapidly to zero as \tpy $\rightarrow$\tc  with increasing dissipation. 
The instability occurs at $dj/dE=0$, which happens 
at a certain value \tpy=\tstar where $H_{c2}(T')$ (or $\eta(T')$) drops 
sufficiently rapidly. \tp depends explicitly only on the power
density $jE = \eta v^{2}B/\Phi_{0}$ and on quantities that depend on \tp
itself (\tauey , specific heat, etc.). Hence at the instability,
$j^{*}E^{*}$=$v^{*2}\eta(T^{*})B/\Phi_{0}$={\em constant}, which
gives the critical-parameter field dependencies:
\begin{equation} \label{bdeps} 
v^{*} \propto 1/\sqrt{B}, E^{*} \propto \sqrt{B},  
j^{*} \propto 1/\sqrt{B}, \mbox{ and } \rho^{*}\propto B 
\end{equation}
using \estary=\vstary $B$, 
\jstary =\vstary$\eta(T^{*})/\Phi_{0}$, and \rhostary=\estary /\jstary .
This gives $v^{*} \propto 1/\sqrt{B}$ in a natural way, 
consistent with our measurements  
in this regime and in contrast to the
$B$-independent \vstar of the pure LO effect near \tcy .

To derive the complete \je response and absolute values of critical
parameters 
we calculate 
\tp  and insert it into Eq.~\ref{tdep-je}.
We take 
\hcuy(0)=120 T \cite{nakagawa} with the WHH function \cite{whh} for 
interpolation between this \hcuy(0)=120 
and \hcuy(\tcy)=0 
(There is some theoretical controversy regarding the exact form of 
\hcuty; however, empirically,  
direct measurements \cite{nakagawa} of \hcuty , within their
uncertainity,
seem not to depart drastically from the WHH function, and
the exact functional shape does not crucially affect our conclusions.).
The dissipation raises the electronic energy by 
$jE\tau_{\epsilon}$, which is related to \tp by 
\begin{equation} 
\label{tprime} 
jE\tau_{\epsilon} \approx
\frac{E^{2}H_{c2}(T')\tau_{\epsilon}}{\rho_{n}(T_{0})B}
\approx \int\limits_{T_{p}}^{T'} c(T) dT
\approx \int\limits_{T_{0}}^{T'} c(T) dT
\end{equation}
where  $c(T)$ is the electronic specific heat.
To calculate $c(T)$, \ybco is modeled as a layered
$d$-wave superconductor: $\Delta_{\vec{k}}(T) = \Delta_{0}(T)[k_{x}^{2}
- k_{y}^{2}]/k^{2} \approx \Delta_{0}(T) \cos (2\theta)$, taking the BCS
temperature dependence for $\Delta_{0}(T)$ and 
$\Delta_{0}(0)$=$19$ meV from tunneling and infrared measurements
\cite{deltavalue}. Then $c(T)$=$\partial\{\sum_{k} E_{k}
f_{k}\}/\partial T$,
where $f_{k}$=$[\exp(E_{k}/k_{B}T) + 1]^{-1}$ is the Fermi-Dirac
distribution, $E_{k} = \sqrt{\zeta_{k}^{2} + \Delta_{k}^{2}}$, 
$\zeta_{k} = \epsilon_{k} -\mu$, and $\mu$=0.2 eV is the chemical 
potential \cite{chempot}. With the replacement of $\sum_{k}$ by $\int
d\zeta N(0)\int d\theta/2\pi$ (where $N(0)$ is the normal density
of states at $\mu$) this $c(T)$ 
is now inserted into Eq.~\ref{tprime},
which is integrated numerically to obtain \tp and thence 
\je from Eq.~\ref{tdep-je}.

\begin{figure}[h]
\begin{center}
        \leavevmode
        \epsfxsize=\hsize
        \epsfbox{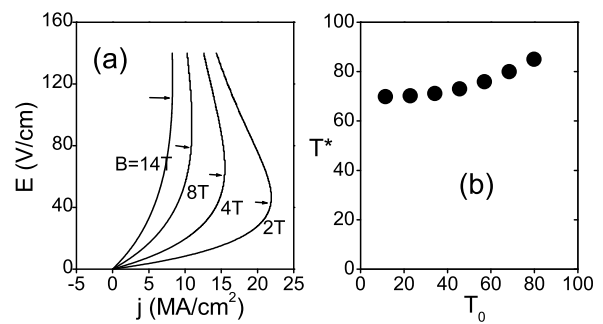}
\end{center}
\vspace{-3.4em}
\caption{\label{fig:epsart} 
Numerical results obtained from solving Eq.~\ref{tprime}, as
described in the text. (a) Theoretical \je curves, at \tooy = 0K. 
The onset of negative slope, indicated by arrows,
marks the instabilities. (b) The 
critical temperature \tstar for different initial temperatures \tooy ; 
\tstar is independent of $B$ and \tauey .}
\label{numerical}
\end{figure}

The numerical results of the above calculation 
are shown in Fig.~\ref{numerical}. The \je curves of 
panel (a) have an infinite slope at the instability 
(\jstary , \estary), indicated by arrows, and then exhibit negative
slope.
This negative sloped portion is experimentally forbidden in a current
biased
measurement and instead will be manifested as a vertical jump in $E$.
The electronic temperature \tp rises from the 
bath value \too to \tstar at the instability. Panel (b) shows the
computed \tstary , which is independent of $B$ as expected 
but has a slight dependence on the bath temperature \tooy .

In order to conveniently scale the experimental curves, the exact 
numerical \je function derived above---and plotted in 
Fig.~\ref{numerical}---can be cast into a mathematically more manageable
form by noting
that the R.H.S. of Eq.~\ref{tprime} can be approximated \cite{mnk-jmk} 
by $jE\tau_{\epsilon} 
\approx  \Delta(T')n_{q}(T') -
\Delta(T_{p})n_{q}(T_{p}) \approx  \Delta(T')n_{q}(T')$. 
$\Delta$ and \nq are connected
through the statistical equations of the previous paragraph. 
Numerically computing the number of quasiparticles excited above the gap 
(taking the anisotropic $d$-wave gap together with a BCS temperature dependence 
as discussed above) we obtain the following $d$-wave generalization 
of the $\Delta - n_{q}$ relationship \cite{mnk-jmk}: 
$(\Delta/\Delta_{0})^{2} = f(n_{q}/n)$, where
$n=2.7\times 10^{21}$ cm$^{-3}$ is the carrier concentration 
and 
$f(x) \simeq 1 - 0.4386 x - 1539 x^{2} + 40381x^{3} - 345217 x^{4}$ 
(despite the appearance of successively increasing
coefficients in $f(x)$, the terms rapidly converge because $x$=\nqy /n$\sim
k_{B}T/E_{F} \sim 0.01$). 
Combining this with the earlier 
$jE\tau_{\epsilon} \approx  \Delta(T')n_{q}(T')$, 
$\eta = j\Phi_{0}/v = j\Phi_{0}B/E$, 
$\eta \approx H_{c2}\Phi_{0}/\rho_{n}$, and $\eta \propto \Delta$, we get
a convenient closed form for the non-linear \je characteristic:
\begin{equation}
\label{jecurve}
j \approx \left(\frac{H_{c2}(T_{0})}{B \rho_{n}(T_{0})}\right)E \sqrt{f(x)}
\end{equation}
with $x = n_{q}/n = 0.0245 \times E^{2}/E^{*2}$, and 
$E^{*} =\sqrt{0.0245\rho_{n}(T_{0})Bn\Delta_{0}/\tau_{\epsilon}
H_{c2}(T_{0})}$ is the value of $E$ at the instability peak. 
We now turn to the experimental details and results. 

The samples are {\em c}-axis oriented epitaxial films of
\ybco on (100) LaAlO$_{3}$ substrates with \tcy 's around 90K and of 
thickness $t \approx 90$ nm. 
Electron-beam 
lithography 
was used to pattern bridges of widths $w \approx$ 2--20 $\mu$m and lengths
$l \approx$ 30--200 $\mu$m. 
Altogether ten samples were studied at 12 temperatures 
(1.6, 2.2, 6, 7, 8, 10, 20, 27, 35, 42, 50, 80 K) and at 13 flux densities
(0.1, 0.2, 0.5, 1, 1.5, 2, 10, 11, 13, 13.5, 13.8, 14, 15.8 T). 
The electrical transport measurements were made with a pulsed 
constant current source, 
preamplifier circuitry, and a digital storage oscilloscope.
The pulse rise times are about 100 ns with a duty cycle of about 1 ppm,
which for the narrowest bridges result 
in effective thermal resistances of order 1 nK.cm$^{3}$/W. 
Note that the $j$ values in the experiment
are an order of magnitude lower than the depairing current 
density \cite{pair} and the applied 
flux densities 
exceed the self field of the current by at least two orders of
magnitude. Further details about the experimental techniques are 
discussed elsewhere \cite{mplb}.

\begin{figure}[h]
\begin{center}
        \leavevmode
        \epsfxsize=\hsize
        \epsfbox{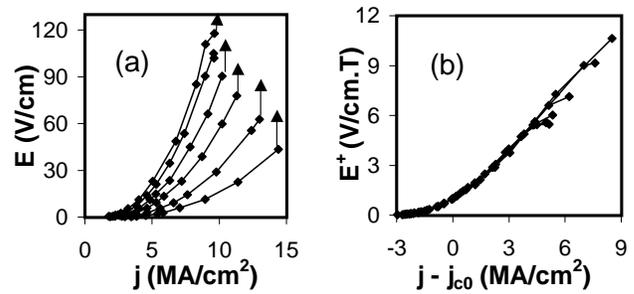}
\end{center}
\vspace{-3em}
\caption{
Experimental current-voltage characteristics. (a) Raw values of $E$
versus $j$ at $T$=20K and applied $B$ values of (from lowest to highest
curve) 3, 5, 8, 11, 13.8, and 15.8 T. The last symbol on each curve
is right at the instability. Slightest further increase of $j$ (entering
the forbidden negative-sloped portion of the theoretical curves of
Fig.~1) 
causes $E$ to make discontinous vertical jumps (arrows).
(b) The same data plotted as $E^{+}=(E/B) \times
\sqrt{f(0.0245E^{2}/E^{*2})}$ 
(as per Eq.~\ref{jecurve})  versus $j -j_{c0}$, where the 
critical depinning current density $j_{c0}$ is defined at $E^{+}=1$
V/cm.T 
(the scaling is not affected by the choice of criterion).}
\label{20k-full-iv}
\end{figure}
Fig.~\ref{20k-full-iv}(a) shows a typical set of experimental \je
curves. The last stable datapoint (\jstary , \estary) 
of each curve is at the tail of each arrow. 
The slightest
further increase of $j>$\jstar causes a drastic vertical jump in $E$
as shown by the arrows
(the voltage pulse jumps off the scale of the oscilloscope upto the
compliance limit of the current source). The jumps show only a 
small hysteresis  $<3$\% of \jstary .
(As expected for a 
current-biased measurement, the break occurs slightly before
the slope has become quite vertical \cite{bezshkl}.)  
Fig.~\ref{20k-full-iv}(b) shows the same data plotted as the R.H.S. of
Eq.~\ref{jecurve} vs  $j -j_{c0}$, the excess current density over the
depinning
value. The data scale well and tend toward homogeneous linearity.   
Note that the collapse implies an excellent proportionality between
$\rho$ and $B$ over the entire range.

\begin{figure}[h]
\begin{center}
        \leavevmode
        \epsfxsize=\hsize
        \epsfbox{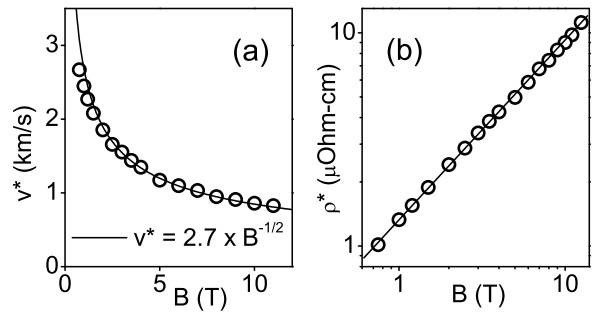}
\end{center}
\vspace{-3em}
\caption{
Variation of critical parameters with flux density.
The measurements were made at $T$$=$$1.6$ K. 
(a) The critical velocity shows a \vstar $\propto 1/\sqrt{B}$ trend
(solid line is a $1/\sqrt{B}$ fit).
 (b) The critical resistivity is
proportional to the flux density (straight line is a guide to the
eye).}
\label{rootbfig}
\end{figure}
Fig.~\ref{rootbfig} shows experimentally measured $B$ dependencies of 
\vstar and \rhostar 
for \tooy=1.6 K, demonstrating excellent
agreement with Eq.~\ref{bdeps} (the other dependencies 
$E^{*} \propto \sqrt{B}$ and $j^{*} \propto 1/\sqrt{B}$ follow from 
$\rho^{*}\propto B$ and $v^{*}\propto 1/\sqrt{B}$). The 
$v^{*}\propto 1/\sqrt{B}$ dependence was found to be ubiquitous for all
of our low-$T$ measurements in ten samples 
(spanning 1.6 K $\leq T \leq$ 50 K and 0.5 T $\leq B \leq$ 15.8 T) 
and has also been seen by Xiao et al. \cite{xiao} at intermediate
temperatures 
(at the lower end of their $T$\aby 60--90 K range of study). 
Note that the excellent linearity between between \rhostar and $B$
demonstrates the independence of $\eta$ on $B$ in this regime; 
then the resistivity is simply 
proportional to the number of vortices and hence $B$.

\begin{figure}[h]
\begin{center}
\end{center}
        \leavevmode
        \epsfxsize=0.6\hsize
        \epsfbox{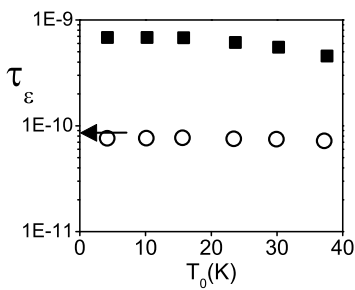}
\vspace{-2em}
\caption{
The energy relaxation time \taue extracted from the measured data using 
the present low-temperature instability model (circles) 
and from LO theory (squares). The leftward arrow represents the
theoretical $T\rightarrow 0$ limiting value of \tauep from   
phonon emission (please see text).}
\label{tauefig}
\end{figure}

The final step in the analysis is to extract \taue from the data. 
Inverting Eq.~\ref{tprime} evaluated at the 
instability (i.e., at \tpy=\tstary ), we get 
$\tau_{\epsilon} \approx \frac{\rho_{n}(T_{0})B}{H_{c2}(T^{*})E^{*2}}
\int\limits_{T_{0}}^{T^{*}} c(T) dT$; the \tstary(\tooy ) function comes
from the model itself (Fig.~\ref{numerical}(b)).
\taue calculated in this way is shown by the circles in 
Fig.~\ref{tauefig}. If the data
are analyzed in the LO framework ($\tau_{\epsilon} \approx
D[14\zeta(3)]^{1/2}B^{2}/(\pi E^{*2})$)
a markedly longer \taue is obtained shown by the squares. For comparison
the arrow shows the theoretical $T$$\rightarrow 0$ 
limiting value estimated from spontaneous emission of
phonons. (Emission of non-equilibrium phonons dominates
{\em ep} scattering by thermal phonons below a cross-over temperature
$T_{x} = m v_{F} c_{s}/k_{B}$$\approx$43 K 
producing an essentially $T$-independent scattering rate 
(since $kT$$\ll$$E_{F}$) of \tauepy 
=$(3\pi \rho_{M} c_{s} \hbar^{2})/(2 C^{2} m k_{F}) 
\approx 8.4 \times 10^{-11}$, where 
$C$$\approx$$6 \times 10^{-13}$ ergs is the coupling between
electron energy and crystal dilation, $c_{s}=4$ km/s 
is the sound velocity, and $\rho_{M}$=6 g/cm$^{3}$ is the mass
density \cite{nistdatabase}; for further discussion see, for example, 
Kittel \cite{kittel}.)

In conclusion, we investigated the low-temperature regime of 
flux motion driven far beyond free flux flow, and  
observed an instability 
under all conditions of fields, and temperatures from $\sim$\tcy/2 down
to essentially $T\approx0$. The nature of this low-temperature
instability
seems to be well
described by a model where the electron gas is heated above the phonon
temperature leading to the generation of quasiparticles and 
loss in viscosity as the vortex core expands and $\Delta$ is reduced.
This scenario is different from the standard LO picture (dominant mainly
near
\tcy ) where the vortex shrinks and quasiparticles leave its vicinity. 
Because the present effect prevails even at temperatures well below 
\tc (where most superconductive devices operate) it becomes an important
consideration in the design of applications where the superconductor
operates in the dissipative regime\cite{fusion}, since the instability
triggers an abrupt rise in dissipation at $j$ values
much below the depairing current density.
Detailed predictions of the model, including field dependencies
of critical parameters
and shapes of \je curves were experimentally verified, and
the effect provides an estimate of the 
$T$$\rightarrow$0 time scale 
for energy exchange between quasiparticles and phonons.

The author gratefully acknowledges useful discussions and other
assistance from
J. M. Knight, B. I. Ivlev, M. Geller, D. K. Christen, J. M. Phillips, R.
P. Huebener, N. Schopohl, J. Blatter, and V. Geshkenbein.
This work was supported by the U. S. Department of Energy 
through grant number DE-FG02-99ER45763.




\begin{references}

\bibitem{lochapter} A. I. Larkin and Yu. N. Ovchinnikov, in {\em
Nonequilibrium Superconductivity}, D. N. Langenberg and A. I. Larkin,
eds. (Elsevier, Amsterdam, 1986), Ch. 11.
\bibitem{bsandtinkhamandclem} J. Bardeen and M. J. Stephen, 
Phys. Rev. {\bf 140}, A1197 (1965); M. Tinkham, \prl {\bf 13}, 804
(1964); and 
J. R. Clem, \prl {\bf 20}, 735 (1968). 
\bibitem{kopninvolovik} N. B. Kopnin and G. E. Volovik, Phys. Rev. Lett.
{\bf 79}, 1377 (1997).
\bibitem{tinkhamtext} A physically intuitive discussion can
be found in Michael Tinkham, {\em Introduction to
Superconductivity}, 2nd Edition (McGraw Hill, New York, 1996).
\bibitem{obs} M. N. Kunchur, D. K. Christen, and J. M. Phillips,
Phys. Rev. Lett. {\bf 70}, 998 (1993).
\bibitem{LO} A. I. Larkin and Yu. N. Ovchinnikov, Zh. Eksp. Teor. Fiz.
{\bf 68}, 1915 (1975) [Sov. Phys. JETP {\bf 41}, 960 (1976)].
\bibitem{klein} W. Klein, R. P. Huebener, S. Gauss, and J. Parisi, J.
Low
Temp. Phys. {\bf 61}, 413 (1985).
\bibitem{loexpt} L. E. Musienko, I. M. Dmitrenko, and V. G. Volotskaya,
Pis'ma Zh. Eksp.Teor. Fiz. {\bf 31}, 603 (1980) [JETP Lett. {\bf 31},
567 (1980)]; and A. V. Samoilov et al.,
\prl {\bf 75}, 4118 (1995).
\bibitem{doett} S. G. Doettinger et al.,
\prl {\bf 73}, 1691 (1994).
\bibitem{xiao} Z. L. Xiao et al., \prb {\bf 53}, 15265 (1996).
\bibitem{eliashberg} G. M. Eliashberg, JETP Lett. {\bf 11}, 114 (1970)
[Sov. Phys. JETP {\bf 34}, 668 (1972)];
and B. I. Ivlev, S. G.
Lisitsyn, and G. M. Eliashberg, J. Low Temp. Phys. {\bf 10}, 449 (1973).
\bibitem{bezshkl} A. I. Bezuglyj and V. A. Shklovskij, 
Physica C {\bf 202}, 234 (1992).
\bibitem{bez} A. I. Bezuglyj, Physica C {\bf 323}, 122 (1999).
\bibitem{volotskaya} V. G. Voltskaya, et al.,  
Sov. J. Low Temp. Phys. {\bf 18}, 683 (1993).
\bibitem{doett2} S. G. Doettinger et al., Physica C {\bf 251}, 285
(1995).
\bibitem{nakagawa} H. Nakagawa, N. Miura, and Y. Enomoto, J. Phys.:
Condens. Matter {\bf 10}, 11571 (1998).
\bibitem{whh} N. R. Werthamer, E. Helfand, and P. C. Hohenberg, Phys.
Rev.
{\bf 147}, 295 (1966).
\bibitem{deltavalue} W. Lanping et al., \prb {\bf 40}, 10954 (1989); and

R. T. Collins et al., \prl {\bf 59}, 704 (1987). 
\bibitem{chempot} C. Poole, H. A. Farach, and R. J. Creswick, 
{\em Superconductivity} (Academic Press, San Diego, 1995).
\bibitem{pair} M. N. Kunchur, D. K. Christen, C. E. Klabunde, and J. M.
Phillips, Phys. Rev. Lett. {\bf 72}, 752 (1994).
\bibitem{mplb} M. N. Kunchur, Mod. Phys. Lett. B. {\bf 9}, 399 (1995).
\bibitem{nistdatabase} NIST 
WebHTS Database, \\http://www.ceramics.nist.gov/srd/hts/htsquery.htm
\bibitem{kittel} C. Kittel, {\em Introduction to Solid State Physics},
seventh edition, (John Wiley and Sons, New York, 1996), appendix J.
\bibitem{fusion} The author has an ongoing project (funded by Lockheed
Martin, Inc.) to explore the high-dissipation regime
for the operation of pulsed superconductive magnets for fusion energy
devices.

\end{references}
\end{document}